# Low threshold optical bistability based on $MoS_2$ in asymmetric Fabry-Perot cavity structure in visible light band

Songqing Tang[1], Mengjiao Ren[1], Zhiheng Li[1], Zhiwei Zheng[1,*] and Leyong Jiang [1,2,†]

[1]*School of Physics and Electronics, Hunan Normal University, Changsha 410081, China*

[2]*Post Moore era laboratory, Hunan Normal University, Changsha 410081, China*

Corresponding Author: *zhengzhiwei@hunnu.edu.cn and †jiangly28@hunnu.edu.cn

This article theoretically proposes a multi-layer Fabry-Perot cavity structure based on nonlinear $MoS_2$, whose cavity is composed of asymmetric photonic crystals. In this structure, we observed a low threshold optical bistability phenomenon on the order of a in the visible light band, which is caused by the large third-order nonlinear conductivity of the bilayer $MoS_2$ and the Fabry-Perot cavity resonance. Research has found that when light is incident from two different directions in an asymmetric Fabry-Perot cavity, the optical bistability exhibits not exactly the same behavior. In addition, we further investigated and found that the optical bistability behavior in this simple multi-layer structure is closely related to parameters such as incident wavelength, Fabry-Perot cavity length, and refractive index of the photonic crystal dielectric. This work provides a new approach for the implementation of low threshold optical bistable devices in the visible light band, which is expected to be applied in nonlinear optical fields such as all optical switches and all optical logic devices.

## 1. Introduction

Optical bistability (OB) refers to a multi value phenomenon similar to a hysteresis loop between the input and output light intensities in an optical system. OB is widely used in fields such as all optical switches [2,3], optical communication [4], biosensing [5], optical storage [6], and all optical logic gates [7]. However, traditional OB materials have weaker nonlinear responses due to their smaller nonlinear coefficients. Therefore, in order to obtain more obvious OB phenomena, OB devices require larger dimensions and stronger incident power during production, which is in line with the requirements of integrated optics for small and low-power consumption, resulting in higher thresholds for OB devices, limiting their applications in micro and nanostructures. In recent years, with the development of micro nano technology, the OB phenomenon in micro nano structures has become an important research direction in the field of OB. In the current context of low response time requirements, low threshold OB has become the core goal pursued by researchers. At present, the ways to reduce the OB threshold can be broadly divided into two categories: the first category is to improve the nonlinear response of the system by searching for optical materials with larger nonlinear coefficients, in order to achieve the goal of reducing the OB threshold; The second type is to achieve the goal of reducing the threshold of OB by constructing specific structures or stimulating specific patterns to achieve local field enhancement effects.

In the search for optical nonlinear materials, in recent years, nonlinear materials such as graphene and Dirac semimetals have stood out among many materials due to their large third-order nonlinear coefficients in the terahertz band [8,9], attracting widespread attention from researchers. Examples include OB in graphene based nanoparticle composite structures [10], OB in graphene based silicon waveguide resonant cavity structures [11], OB in graphene based Kretschmann prism structures [12], OB in Dirac semimetal based photonic crystal heterostructures [13], and OB in Dirac semimetal based photonic crystal Fabry Perot structures [14]. Researchers have focused on micro nano structures such as Fabry Perot cavities [15], prism coupling [16], photonic crystals [17], grating structures [18], as well as surface plasmons [19], topological edge states [20], optical Tamm states [21], and continuously bound states [22] to enhance and reduce the OB threshold by constructing specific structures and exciting specific modes. Based on these structures and modes, researchers have conducted extensive research in the OB field. Especially recently, Li et al. proposed a graphene based nanoarray structure and utilized it to achieve low threshold OB phenomena in the mid infrared band under the excitation of surface plasmon resonance mode [23]. He et al. proposed an asymmetric one-dimensional photonic crystal structure containing Weyl semimetals, which achieved low threshold optical bistable absorption in the terahertz band by utilizing the local field enhancement effect caused by Tamm plasmon excitation [24]. Kim et al. proposed a one-dimensional Si photonic crystal structure with $SiO_2$ as the background material, which induced the generation of bound states. They demonstrated the bistability of the electrical wavelength under room temperature and thermally induced conditions, achieving a low threshold bistable phenomenon [25]. However, in order to achieve greater third-order nonlinear conductivity, more OB implementation approaches are still mainly focused on the terahertz band. At present, there are still not many studies on OB at higher frequencies than terahertz, and the search for nonlinear materials remains a challenge, especially in the visible light band.

Recently, $MoS_2$ has attracted the attention of researchers due to its excellent optoelectronic properties. It is one of the transition metal dihalide compounds, with a structure similar to graphene. It is a type of graphene material with many similar physical and chemical properties [26], and is widely used in optical fields such as photodetectors [27], optical storage devices [28], and optical transistors [29]. It is worth mentioning that in recent years, research has found that when $MoS_2$ bilayer exists, it can act as a Kerr medium as a whole, with a large third-order nonlinear conductivity in the visible light band [30], which provides nonlinear conditions for the implementation of low threshold OB in the visible light band. Therefore, we optimistically infer that, based on the nonlinear conditions provided by $MoS_2$, it is possible to achieve controllable low threshold OB in the visible light band by combining specific structures and modes.

In this article, we design an asymmetric photonic crystal Fabry-Perot cavity structure based on nonlinear $MoS_2$, which can achieve low threshold OB phenomenon in the visible light band. We found that the behavior of OB changes to a certain extent when light is incident from two different directions, as well as when the incident wavelength, incident angle, Fabry-Perot cavity length, and dielectric refractive index of the photonic crystal are changed. This provides a new approach for tunable OB implementation in the visible light band. Meanwhile, the OB structure is simple and easy to prepare, and it has the potential to be applied in nonlinear optical fields such as all optical switches and all optical logic devices.

## 2. Theoretical Model and Method

We consider an asymmetric photonic crystal Fabry-Perot cavity structure based on nonlinear $MoS_2$. In this structure, the bilayer $MoS_2$ is embedded between two photonic crystal PhCs, as shown in Fig 1. These two PhCs are composed of a periodic alternating arrangement of medium A and medium B, where medium A is Si and the refractive index is set to 2.82 [31], $n_{a1} = n_{a2} = n_a = 2.82$, medium A is Si and the refractive index is set to 1.5[32], $n_{b1} = n_{b2} = n_b = 1.5$。 The thicknesses of medium A and medium B respectively satisfy $d_{a1} = d_{a1} = d_a = \lambda_c/4n_a$, $d_{b1} = d_{b1} = d_b = \lambda_c/4n_b$ 。 The period of PhC1 is set to $T_1=5$, and the period of PhC2 is set to $T_2=6$. $\lambda_c$ is the central wavelength, $\lambda_c = 579.1$ nm. L is the cavity length of the Fabry-Perot cavity, set to L=546.5 nm, and the incidence angle is $\theta = 53°$. The thickness of each layer of $MoS_2$ is 0.65 nm, and under random phase conditions without considering the effect of external magnetic fields, the dielectric constant of $MoS_2$ can be expressed as [33,34]：

$$\varepsilon^{(1)}_{(\omega)} = \varepsilon_{\infty 2} + \sum_{i=0}^{5} \frac{a_i \omega_p^2}{\omega_i^2 - \omega^2 - ib_i\omega} - \frac{2\alpha}{\sqrt{\pi}} DawsonF(\frac{\mu - \hbar\omega}{\sqrt{2}\sigma}) + \alpha \exp(-\frac{(\hbar\omega - \mu)^2}{2\sigma^2}), \quad (1)$$

where $\alpha = 23.234$, $\mu = 2.7723$ eV, $\sigma = 0.3089$ eV. $\varepsilon_{\infty 2} = 4.44$ is the DC dielectric constant, $a_i$ represents the oscillator strength, $b_i$ represents the damping coefficient, $\omega_i$ represents the response frequency, and $\omega_p = 7 \times 10^{12}$ rad/s represents the plasma frequency [34]. In addition, the first-order linear conductivity, third-order nonlinear conductivity, and total conductivity of MoS$_2$ can be expressed as [30]:

$$\begin{cases} \sigma_{(\omega)}^{(1)} = \dfrac{\Lambda k_0}{i\eta_0}(\varepsilon_{(\omega)}^{(1)} - 1), & (2\text{-}1) \\[6pt] \sigma_{(\omega)}^{(3)} = \dfrac{2\Lambda k_0}{i\eta_0}\dfrac{3m_e \omega_0^2 \varepsilon^3}{d^2 N^3 e^4}\left[(\varepsilon_{(\omega)}^{(1)} - 1)^3\right]\left(\varepsilon_{(-\omega)}^{(1)} - 1\right), & (2\text{-}2) \\[6pt] \sigma = \sigma_{(\omega)}^{(1)} + \sigma_{(\omega)}^{(3)}\left|E_{11y}(z = 5d_a + 5db + L)\right|^2, & (2\text{-}3) \end{cases}$$

where $\eta_0$=377 Ω represents vacuum resistivity, $k_0 = \omega/c$ represents wave vector in vacuum, and $\omega_0 = 4.51 \times 10^{15}$ rad/s represents resonant frequency. Here $d$ = 3Å represents the size of atoms and $N$=10$^{28}$ m$^{-3}$ represents the density of atomic numbers. For the convenience of calculation, we will only consider the TE polarization mode in the subsequent calculations. We set the propagation direction of electromagnetic waves as the z-direction, and the plane where MoS$_2$ is located is the xoy plane. According to Maxwell's equations, the electromagnetic field in the leftmost incident medium (air) can be represented as:

$$\begin{cases} E_{iy} = E_i e^{ik_{iz}z} e^{ik_x x} + E_r e^{-ik_{iz}z} e^{ik_x x}, & (3\text{-}1) \\ H_{ix} = -\dfrac{k_{iz}}{\mu_0 \omega} E_i e^{ik_{iz}z} e^{ik_x x} + \dfrac{k_{iz}}{\mu_0 \omega} E_r e^{-ik_{iz}z} e^{ik_x x}, & (3\text{-}2) \\ H_{iz} = \dfrac{k_x}{\mu_0 \omega} E_i e^{ik_{iz}z} e^{ik_x x} + \dfrac{k_x}{\mu_0 \omega} E_r e^{-ik_{iz}z} e^{ik_x x}. & (3\text{-}3) \end{cases}$$

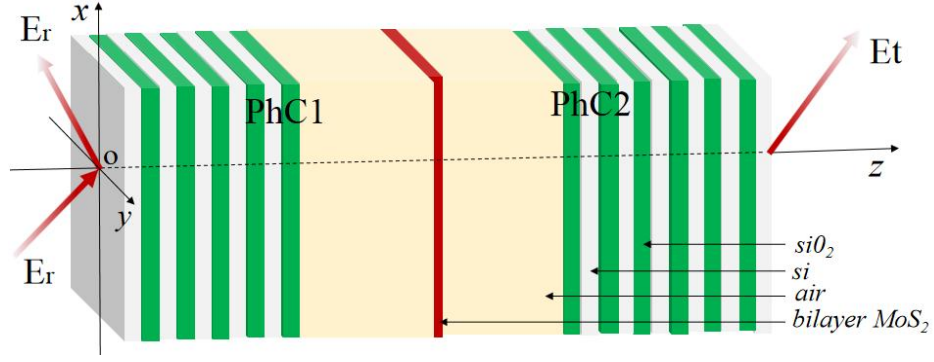

Fig.1. Schematic diagram of the Fabry Perot cavity structure in photonic crystals

In the first dielectric, the electromagnetic field can be represented as:

$$\begin{cases} E_{1y} = F_1 e^{ik_{1z}(z-d_a)} e^{ik_x x} + B_1 e^{-ik_{1z}(z-d_a)} e^{ik_x x}, & (4\text{-}1) \\ H_{1x} = -\dfrac{k_{1z}}{\mu_0 w} F_1 e^{ik_{1z}(z-d_a)} e^{ik_x x} + \dfrac{k_{1z}}{\mu_0 w} B_1 e^{-ik_{1z}(z-d_a)} e^{ik_x x}, & (4\text{-}2) \\ H_{1z} = \dfrac{k_{1z}}{\mu_0 w} F_1 e^{ik_{1z}(z-d_a)} e^{ik_x x} - \dfrac{k_{1z}}{\mu_0 w} B_1 e^{-ik_{1z}(z-d_a)} e^{ik_x x}. & (4\text{-}3) \end{cases}$$

Afterwards, the electromagnetic field in each intermediate layer of medium can be represented as:

$$\begin{cases} E_{my} = F_m e^{ik_{jz}(z-z_0)} e^{ik_x x} + B_m e^{-ik_{jz}(z-z_0)} e^{ik_x x}, & (5\text{-}1) \\ H_{mx} = -\dfrac{k_{jz}}{u_0 \omega} F_m e^{ik_{jz}(z-z_0)} e^{ik_x x} + \dfrac{k_{jz}}{u_0 \omega} B_m e^{-ik_{jz}(z-z_0)} e^{ik_x x}, & (5\text{-}2) \\ H_{mz} = \dfrac{k_x}{u_0 \omega} F_m e^{ik_{jz}(z-z_0)} e^{ik_x x} + \dfrac{k_x}{u_0 \omega} B_m e^{-ik_{jz}(z-z_0)} e^{ik_x x}. & (5\text{-}3) \end{cases}$$

The electromagnetic field in the rightmost emitting medium can be expressed as:

$$\begin{cases} E_{sy} = E_t e^{ik_{sz}(z-10d_a-10db-2L)} e^{ik_x x}, & (6\text{-}1) \\ H_{sx} = -\dfrac{k_{0z}}{\mu_0 w} E_t e^{ik_{sz}(z-10d_a-10db-2L)} e^{ik_x x}, & (6\text{-}2) \\ H_{sz} = \dfrac{k_{0z}}{\mu_0 w} E_t e^{ik_{sz}(z-10d_a-10db-2L)} e^{ik_x x}. & (6\text{-}3) \end{cases}$$

where $k_{0z} = k_0 cos(\theta)$, $k_{0x} = k_0 sin(\theta)$, $u_0$ represents vacuum magnetic permeability, $E_i$、$E_r$、$E_t$ represent incident electric field, reflected electric field, and transmitted electric field, respectively. $F_n$ represents the amplitude of electric field propagation along the positive z-axis, while $B_n$ represents the amplitude of electric field propagation along the opposite z-axis. When electromagnetic waves propagate at the dielectric interface $z = \alpha d_a + \beta db + \gamma L$, the electric field is continuous while the magnetic field is discontinuous, and its propagation satisfies the boundary conditions:

$$\begin{cases} E_{my}(\alpha d_a+\beta db+\gamma L) = E_{(m+1)y}(\alpha d_a+\beta db+\gamma L) & (7\text{-}1) \\ H_{mx}(\alpha d_a+\beta db+\gamma L) = H_{(m+1)x}(\alpha d_a+\beta db+\gamma L) & (7\text{-}2) \end{cases},$$

Specifically, at the interface $z=5d_a+5db+L$, due to the presence of a thin layer of MoS₂, the boundary conditions differ, specifically:

$$E_{11x}(z = 5d_a+5db+L) = E_{12x}(z = 5d_a+5db+L), \tag{8}$$

$$H_{11y}(z = 5d_a+5db+L) - H_{12y}(z = 5d_a+5db+L) = \sigma E_{11x}(z = 5d_a+5db+L), \tag{9}$$

Based on the above formula, we can ultimately obtain the relationship between the transmission electric field $E_t$ and the transmittance, as well as the incident electric field $E_i$. Subsequently, by adjusting appropriate structural parameters, nonlinear optical bistability phenomena can be observed.

## 3. Results and Discussions

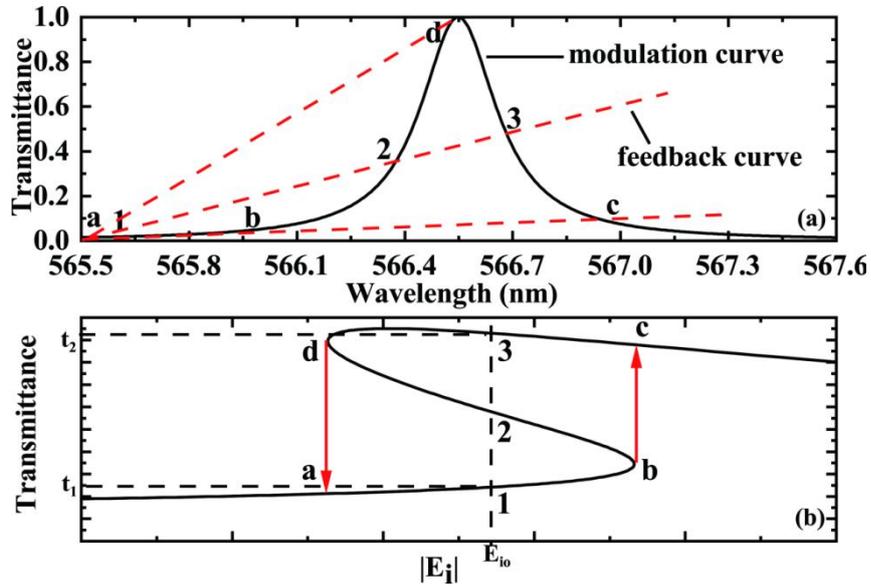

Fig. 2. The relationship between transmittance and incident wavelength. (b) Diagram of the variation of transmitted electric field with incident electric field.

In this section, we have conducted a targeted discussion on the OB phenomenon in Fabry-Perot cavities. Firstly, we calculated the transmittance curve of the entire structure using the transfer matrix method, considering only the linear case of $MoS_2$, as shown in Fig. 2(a). In this figure, the reflectance curve corresponding to the solid line is called the modulation curve, and the three dashed lines are called the feedback curve. The slope of the feedback curve is directly proportional to the reciprocal of the incident light intensity. As shown in the figure, there is a clear transmission peak at $\lambda = 566.5$ nm in the modulation curve, which is caused by the cavity resonance of the Fabry-Perot cavity [35]. When the Fabry-Perot cavity resonates, it will lead to the generation of localized field enhancement effect. Considering this effect and not considering specific wavelengths, we replaced linear $MoS_2$ with nonlinear $MoS_2$ and obtained the relationship curve between transmittance and incident electric field, which is shaped like an 'S' curve, as shown in Fig. 2(b). The formation of this S-shaped curve can be well explained by the relationship between the modulation curve and feedback curve in Fig. 2(a). In Fig. 2(a), it is not difficult to see from the figure that there are three intersections between the modulation curve and the feedback curve between the straight lines ad and bc, that is, there are three different transmittance values under the same incident light intensity. This means that one incident light intensity corresponds to three different transmitted light intensities. If the incident electric field is gradually increased, according to the feedback relationship [35], the slope of the feedback curve gradually decreases, and the intersection points of the straight line and the curve are a-1-b-c. If the electric field is gradually reduced, the slope of the feedback curve gradually increases, and the intersection points are c-3-d-a in sequence. This is in good agreement with the

S-shaped curve in Fig. 2(b). On this curve, when it is very small, the Transmittance slowly increases with the increase of the incident electric field, corresponding to the a-1-b process, which is a stable state. However, as the incident electric field gradually increases until $|E_i|_{down}$, the Transmittance will jump to another stable state c-3-d. In this state, although the Transmittance continues to decrease, it will not immediately return to the first stable state.On the contrary, when it is very large, the system is in the second stable state, and the Transmittance decreases as it decreases. When it decreases to $|E_i|_{up}$, transmittance will jump from the second stable state c-3-d to the first stable state. It is worth noting that there is also an unstable state b-2-d in between these two stable states, in which the increase in Transmittance accompanied by it shows an unstable increase, and this process cannot be observed in the experiment. For the a-b and c-d regions that can be observed in the experiment, there are two corresponding reflectivity points in this region, which leads to the formation of two jumping discontinuities, b and d, and thus promotes the formation of hysteresis loops. The hysteresis width is $\Delta|E_i|=|E_i|_{down}-|E_i|_{up}$, which is a classic OB phenomenon. In the formation process of this classic OB phenomenon, the third-order nonlinear conductivity of double-layer $MoS_2$ and the localized field enhancement effect caused by Fabry-Perot cavity resonance play a crucial role. Therefore, in the subsequent control work of OB, we mainly consider two aspects: the influence of parameter changes on the third-order nonlinear conductivity of $MoS_2$ and the distribution of local field enhancement.

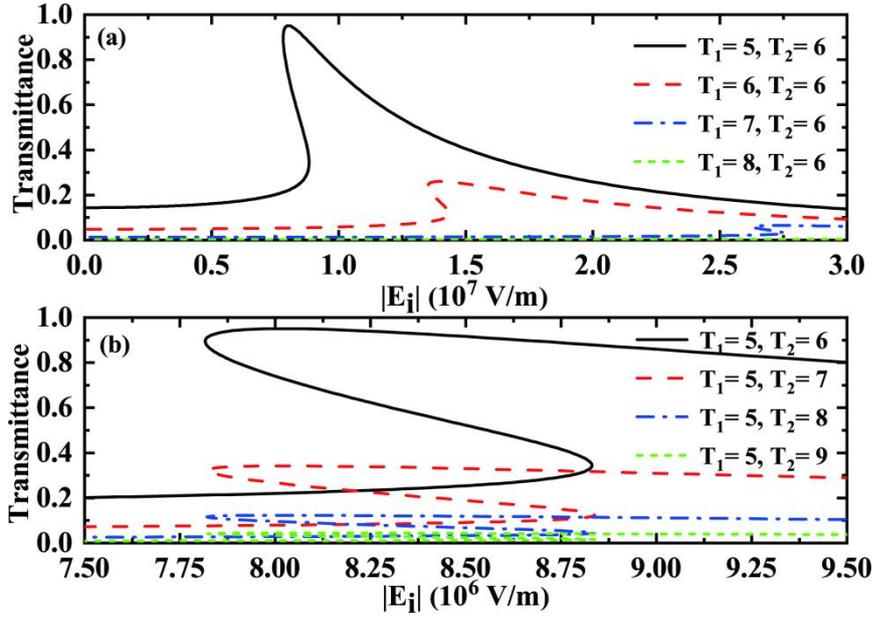

Fig. 3.(a) The relationship between transmittance and the incident electric field when the electric field is incident from left to right. (b)The relationship between transmittance and the incident electric field when the electric field is incident from right to left.

Next, we used the method in the second part to calculate the relationship curves between transmittance and incident electric field when light is incident from left to right and from right to left, respectively. The remaining parameters are consistent with Figure 1, and the results are shown in Fig. 3(a) and Fig. 3(b). It is not difficult to see from the figure that when light is incident from two different directions, OB exhibits different behaviors, with significant changes in its upper and lower threshold values, as shown by the solid lines in Fig. 3(a) and Fig. 3(b). When light is incident from left to right, the upper limit threshold of OB is $|E_i|_{up} = 7.81 \times 10^6$ V/m, the lower limit threshold is $|E_i|_{down} = 8.83 \times 10^6$ V/m, and the threshold width is $\Delta|E_i| = 1.02 \times 10^6$ V/m. When light is incident from right to left, the upper limit threshold of OB is $|E_i|_{up} = 1.33 \times 10^7$ V/m, the lower limit threshold is $|E_i|_{down} = 1.41 \times 10^7$ V/m, and the threshold width is $\Delta|E_i| = 0.08 \times 10^7$ V/m. After discovering that OB exhibits different characteristics when the electromagnetic field is incident from different directions, we further discussed the impact of changes in the photonic crystal period $T_1$ and $T_2$ on OB behavior when incident from the left and right directions. As shown in Fig. 3 (a), when light is incident from left to right, the fixed period of PhC2 is $T_2=6$, while the period of PhC1 $T_1$ is changed. It is not difficult to observe that as $T_1$ increases, the threshold of OB continues to increase, while the peak transmittance of the structure decreases significantly. when $T_1 = 5$, the threshold of OB appears on the order of $10^6$ V/m, and the maximum value of transmittance was 0.95. However, gradually increasing the period of PhC1, when $T_1 = 6$, the threshold of OB appears on the order of $10^7$ V/m, and the maximum value of transmittance significantly decreases to 0.25. Continue to increase $T_1$, and when $T_1 = 8$, the transmittance curve almost shows a straight line shape. Similarly, in Fig. 3(b), light is incident from right to left, with the fixed period $T_1=5$ of PhC1 remaining unchanged, while the period $T_2$ of PhC2 is changed. It can be clearly observed that with the continuous increase of $T_2$, the upper and lower thresholds, as well as the threshold width of OB, have hardly changed, but the overall transmittance of the structure has rapidly decreased. When it continues to increase to $T_2 = 9$, the maximum and minimum values of the OB transmittance curve almost stick together. Therefore, based on the above results analysis, it can be concluded that by selecting the correct direction of light incidence and setting the $T_1$ and $T_1$ reasonably, we can produce the low threshold OB device we need.

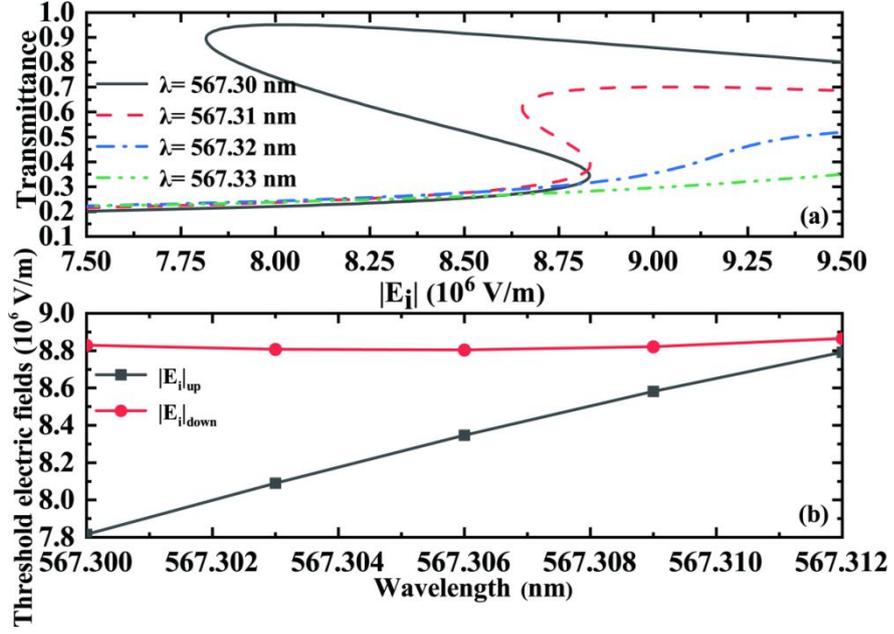

Fig. 4.(a) The relationship between transmittance and incident electric field at different incident wavelengths. (b) The upper and lower thresholds of OB at different incident wavelengths.

Subsequently, we analyzed the relationship curve between the transmitted electric field and the incident electric field at different incident wavelengths, and the remaining parameters remained consistent with Fig. 1, as shown in Fig. 4(a). It is not difficult to see from the graph that the behavior of OB will undergo significant changes when the values of the incident wavelength are different. As the incident wavelength gradually decreases, the upper threshold of OB $|E_i|_{up}$ rapidly increases while the lower threshold $|E_i|_{down}$ remains almost unchanged, resulting in a narrowing of the threshold width. In order to more intuitively describe the impact of incident wavelength on OB threshold, we further plotted the upper and lower threshold values of OB at different incident wavelengths, as shown in Fig. 4(b). When the incident wavelength is set to $\lambda=567.303$ nm, the upper thresholds of OB is $|E_i|_{up}=8.09\times10^6$ V/m, and the lower thresholds of OB is $|E_i|_{down}=8.81\times10^6$ V/m, the threshold width is $\Delta |E_i| = 0.72 \times 10^6$ V/m. When the incident wavelength is set to $\lambda = 567.306$ nm, the upper thresholds of OB is $|E_i|_{up} = 8.35 \times 10^6$ V/m, and the lower thresholds of OB is $|E_i|_{down} = 8.81 \times 10^6$ V/m, and the threshold width is $\Delta |E_i| = 0.46 \times 10^6$ V/m. When the incident wavelength is set to $\lambda = 567.309$ nm, the upper thresholds of OB is $|E_i|_{up} = 8.58 \times 10^6$ V/m, and the lower thresholds of OB is $|E_i|_{down} = 8.82 \times 10^6$ V/m, and the threshold width is $\Delta |E_i| = 0.24 \times 10^6$ V/m. When the threshold width continues to increase, the hysteresis width of OB disappears. Obviously, changes in the incident wavelength can cause changes in the upper and lower thresholds of OB, and their changes can also have a certain regulatory effect on the threshold width of OB. The reason why adjusting the incident wavelength can regulate the behavior of OB is that changing the incident wavelength can change the wave vector in vacuum, thereby changing the overall conductivity of $MoS_2$ and thus changing the behavior of OB, as shown in formula (2). After considering the influence of adjusting the incident wavelength on the conductivity of $MoS_2$ and thus on the behavior of OB, we will analyze the impact of changes in structural parameters on the behavior of OB in the following section.

Subsequently, in this section, we discussed the influence of the cavity length L of the Fabry-Perot cavity on the OB behavior. The remaining parameters are consistent with Fig. 1, and the results are shown in Fig. 5(a) and Fig. 5 (b). As one of the important components of the Fabry-Perot cavity structure, the length variation of cavity length L sensitively affects the hysteresis behavior of OB. It is not difficult to see from the figure that as the length of the cavity gradually increases, the upper limit threshold of OB gradually decreases, while the lower limit threshold remains almost unchanged, which leads to an increase in the width of OB threshold. For example, when L=546.48 nm, the upper thresholds of OB is $|E_i|_{up}=8.57\times10^6$ V/m, and the lower thresholds of OB is $|E_i|_{down}=8.79\times10^6$ V/m, and the threshold width is $\Delta|E_i|=0.22\times10^6$ V/m. When L=546.49 nm, the upper thresholds of OB is $|E_i|_{up}=8.21\times10^6$ V/m, and the lower thresholds of OB is $|E_i|_{down}=8.78\times10^6$ V/m, and the threshold width is $\Delta|E_i|=0.57\times10^6$ V/m. When L=546.50 nm, the upper thresholds of OB is $|E_i|_{up}=7.81\times10^6$ V/m and the lower thresholds of OB is $|E_i|_{down}=8.83\times10^6$ V/m, and the threshold width is $\Delta|E_i|=1.02\times10^6$ V/m. The above results indicate that changes in the cavity length of the Fabry-Perot cavity can significantly affect the behavior of OB. The reason why changing the cavity length of the Fabry-Perot cavity can regulate the behavior of OB is because the subtle change in the cavity length of the Fabry-Perot cavity leads to a certain degree of change in the resonant wavelength of the cavity, and a change in the position of the resonant wavelength of the cavity can cause a change in OB behavior. Therefore, the change in Fabry-Perot cavity length can be regarded as an important OB regulation method.

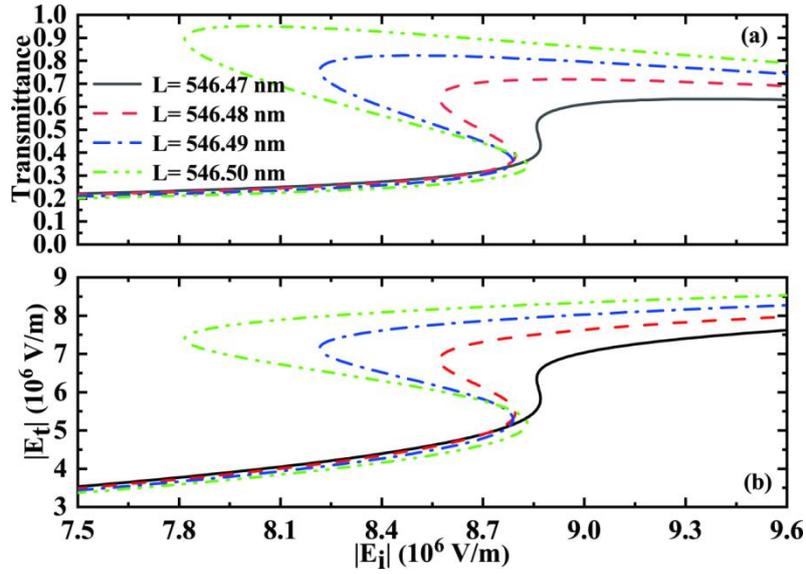

Fig. 5. The relationship curves between (a) transmittance and (b) transmitted electric field and incident electric field at different cavity lengths

Finally, we discussed the influence of the refractive index variation of dielectric B in the Fabry Perot cavity structure of photonic crystals on the OB behavior. The remaining parameters are consistent with Fig. 1, and the results are shown in Fig. 6 (a) and Fig.6(b). It is not difficult to observe that with $n_b$ slow increase, the upper and lower thresholds of OB are also gradually increasing. However, the rate of increase in the lower threshold of OB is greater than that of the upper threshold, which leads to an increase in the width of the OB threshold. For example, when the refractive index of dielectric B is $n_b=1.500$, the upper thresholds of OB is set to $|E_i|_{up}=8.83\times10^6$ V/m, and the lower thresholds of OB is set to $|E_i|_{down}=10.17\times10^6$ V/m, and the threshold width is set to $\Delta|E_i|=1.34\times10^6$ V/m. When the refractive index of dielectric B is $n_b=1.501$, the upper thresholds of OB is set to $|E_i|_{up}=9.88\times10^6$ V/m, and the lower thresholds of OB is set to $|E_i|_{down}=11.56\times10^6$ V/m, and the threshold width is set to $\Delta|E_i|=1.68\times10^6$ V/m. When the refractive index of dielectric B is $n_L=1.503$, the upper thresholds of OB is set to $|E_i|_{up}=10.99\times10^6$ V/m and the lower thresholds of OB is set to $|E_i|_{down}=11.56\times10^6$ V/m, and the threshold width is set to $\Delta|E_i|=3.03\times10^6$ V/m. The above results indicate that changing the refractive index of dielectric B can also achieve controllable adjustment of OB behavior, and it can also be used as a means of manufacturing and designing tunable OB devices.

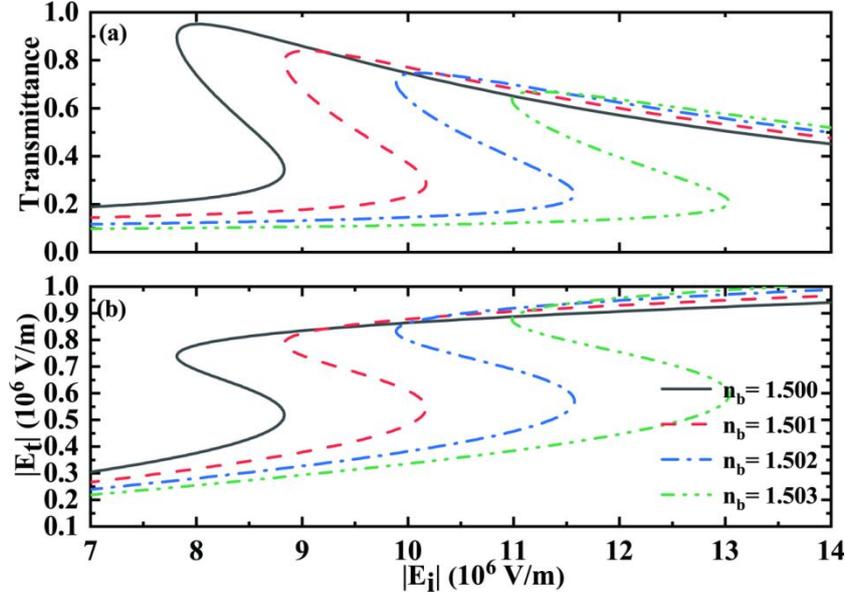

Fig. 6. The relationship curves between (a) transmittance and (b) transmission electric field and incident electric field under different refractive indices of Fabry-Perot cavity dielectric B in photonic crystals.

## 4. Conclusions

In summary, we propose an asymmetric photonic crystal Fabry-Perot cavity structure based on $MoS_2$. Through parameter optimization, we utilized this structure to achieve a low threshold OB phenomenon on the order of $10^6$ V/m in the visible light band. We found that when light is incident from two different directions, OB exhibits different behaviors. Further research has found that the behavior of this OB is also influenced by changes in parameters such as incident wavelength, incident angle, Fabry Perot cavity length, and refractive index of photonic crystal dielectrics. This provides an important approach for the study of tunable OB in the visible light band. Meanwhile, the OB research scheme has a simple structure and is easy to prepare, which has the potential to be applied in nonlinear optical fields such as all optical switches and all closed logic gates.

## Acknowledgments

This work is partially supported by the National Natural Science Foundation of China (Grant No. 62375084).